\documentclass[twocolumn,showpacs,preprintnumbers,amsmath,amssymb,showkeys]{revtex4}

 \usepackage{epsfig}
\usepackage{amssymb}
\usepackage{bm}
\newcommand{\ba}{\begin{eqnarray}}
\newcommand{\ea}{\end{eqnarray}}
\newcommand{\be}{\begin{equation}}
\newcommand{\ee}{\end{equation}}
\newcommand{\bd}{\begin{displaymath}}
\newcommand{\ed}{\end{displaymath}}

\newcommand{\w}{\varpi}

\newcommand{\m}{\dot{M}_+}

\def\lesssim{\mathrel{\hbox{\rlap{\hbox{\lower4pt\hbox{$\sim$}}}\hbox{$<$}}}}
\def\gtrsim{\mathrel{\hbox{\rlap{\hbox{\lower4pt\hbox{$\sim$}}}\hbox{$>$}}}}

\begin{document}
\setcounter{page}{1}
\title{Poynting Flux out of Rotating Black Hole and Accretion Flow through Force-Free
Magnetosphere\footnote{This paper was presented at APCTP Winter School on Black Hole
Astrophysics 2006.}}
\author{Hyun Kyu  \surname{Lee}}
\email{hyunkyu@hanyang.ac.kr}
\affiliation{Department of Physics, Hanyang University, Seoul
133-791}
\begin{abstract}
 The basic features of the Poynting flux  from the horizon
and the ergosphere of a black hole and from the accreting flow
onto a black hole are discussed for the force-free magnetosphere.
The accretion flow dominated by the Poynting flux is discussed
and the possible Poynting flux from the equatorial plane inside
the ergosphere is discussed.
\end{abstract}

\pacs{97.10.Gz, 97.60.Lf}

\keywords{Poynting flux, Black hole, Accretion flow, Force-free
magnetosphere}

\maketitle

\section{INTRODUCTION}

The studies of the explosive processes like SNe and GRBs  have
been revealing the evidences of compact objects such as  neutron
star and/or black hole at the center of explosive events. And it
is likely that massive black holes are responsible for the
activities of AGN. The existence of a black hole  leads to the
idea that the energy sources for powering violent astrophysical
phenomena like AGN and GRB  are likely to be  the gravitational
binding energy of the accreting material onto a black hole and the
rotational energy of a black hole itself\cite{BZ,hkl, lovelace,
blandford}. It has been also expected that the strong and large
scale magnetic field configuration can be formed  around the
compact objects at the centers,  particularly for the late stage
of the core-collapses of rapidly rotating and strongly magnetized
stars. It is then natural to suppose a simple physical situation
in which the energy of the accreting material and the rotational
energy of a black hole can be extracted out through the magnetic
field lines in the form of the Poynting flux.

The Poynting flux has been considered to be a viable route  to
transport the energy and angular momentum along the magnetic field
lines and  one of the efficient ways to tap the  black hole's
rotational energy using  magnetic field lines anchored on the
horizon\cite{BZ} supported by the external current. The relevance
of using  the Poynting flux  in describing  the powerful and
highly collimated astrophysical jets observed in AGN and quasars
has been suggested long time ago \cite{lovelace,blandford} and
many interesting works have been developed.  One of the
characteristics of the  Poynting flux is that it carries very
little baryonic component compared to the hydrodynamic flow. This
property of Poynting flux is found to be consistent with the
required property for powering GRB\cite{piran}.  Recently
numerical simulations also have begun to demonstrate the
electromagnetic energy extraction \cite{Hawley,mg,kom2} from a
black hole surrounded by the force-free magnetosphere, which  can
be established when the electromagnetic field is strong enough
with sufficient charged particles for space-charges and currents
and the inertia of the plasma can be ignored. The Poynting flux in
a system of black hole-accretion disk has also been studied in
connection with GRBs\cite{lwb,lbw,li1} and it is found that the
evolution of the system   largely depends on the Poynting outflow
from the disk\cite{lk,li,wang, uzdensky}.

In this work,  the nature of the Poynting flux  from the horizon
and the ergosphere of a black hole and from the accreting flow
onto a black hole will be discussed.  To make the discussion
simpler  the environment through which the Poynting flux carries
energy out is assumed to be force-free with steady and
axisymmetric configuration and the accreting flow is assumed to be
confined on the equatorial plane. In section II, the force-free
magnetosphere is introduced together with the consistent
constraint and the basic features of the Poynting flux is
discussed in section III. In section IV, the accretion flow
dominated by the Poynting flux is discussed briefly and the
possible Poynting flux from the equatorial plane inside the
ergosphere is discussed in section V.

\section{Force-free Magnetosphere in Kerr Background}

The force-free magnetosphere  can be established when the
electromagnetic field is strong enough with sufficient charged
particles for space-charges and currents and the inertia of the
plasma can be ignored. The role of the plasma in the force-free
magnetosphere is to provide the charge and current sources for the
field. The mathematical expression for the force-free limit is
given by  \ba F_{\mu\nu} J^\nu =0,
        \label{ff}\ea
where $J^\nu$ is the electromagnetic current density.  In terms of
the physical quantity defined by FIDO, it can be written as \ba
\rho \vec{E} + \vec{J} \times \vec{B} =0, \label{ff1} \ea from
which one can also observe that \ba \vec{E} \cdot
\vec{B}=0,\label{ff2} \ea
 which  implies
that the electric potential is conserved along the magnetic field
lines.  It should be noted that the force-free condition is not
identical to the ideal MHD condition, \ba F_{\mu\nu} u^\nu =0,
\label{mhd}\ea  or equivalently \ba \vec{E} + \vec{v} \times
\vec{B} =0, \label{mhd1} \ea where $u^\nu$ is the four velocity of
the plasma.

The force-free requirement  should be consistent with the
condition that  there is no local frame in which a pure electric
field can be seen\cite{znajek,kom1}.  In other words, the
invariant \ba B^2-E^2=-\frac{1}{2}F^{\mu\nu}F_{\mu\nu} \geq 0,
\label{be0}\ea   has to be always checked.   On the other hand,
when one can find a region where Eq.(\ref{be0}) can not be
consistent  it is very likely that it is the signature of a
break-down of force-free condition. Hence it can be a  very useful
method to find out the non-force-free region (most likely a part
of the region) even without knowing the detailed knowledge of the
plasma around.

In the steady and axisymmetric case around the Kerr black hole, we
get from the the force-free condition, Eq.(\ref{ff}),
        \ba \vec{E}=-{\bar{\omega}\over{\alpha}} (\Omega_F+\beta) ~
        e^{\hat{\phi}} \times \vec{B}^p \label{fff2}. \ea
$\Omega_F$, which is  defined by $dA_0 = -\Omega_F dA_{\phi}$,
 is the angular velocity of the magnetic
surface that rotates rigidly in an axisymetric and stationary
state \cite{TPM}.  Throughout this work, the background metric is
assumed to be the Kerr metric\cite{kerr}.  Using the
 Boyer-Lindquist coordinates\cite{bl} in the
 natural unit $G=c=1$,  the non vanishing components
 are given by
\begin{eqnarray}
g_{00}= -(\alpha^2 -\varpi^2 \beta^2), ~~
g_{0\phi}=g_{\phi0}=\varpi^2\beta, ~~ \nonumber \\
g_{rr}=\frac{\rho^2}{\Delta} , ~~ g_{\theta \theta}= \rho^2, ~~
g_{\phi \phi}=\varpi^2, \label{kerr}
\end{eqnarray}
where
\begin{eqnarray}
\alpha &=& \frac{\rho \sqrt{\Delta}}{\Sigma}, ~~
\beta=-\frac{2aMr}{\Sigma^2}, ~~ \tilde{\omega} =
\frac{\Sigma}{\rho} \sin \theta, \nonumber \\~~\Delta &=& r^2 +
a^2 -2Mr, ~~ \rho^2 = r^2 + a^2 \cos^2 \theta, ~~\nonumber \\
\Sigma^2 &=& (r^2 + a^2)^2 - a^2 \Delta
\sin^2\theta.\label{tomega}
 \end{eqnarray}

Using Eq.(\ref{fff2}), the invariant $B^2-E^2$ can be written as
\ba B^2-E^2 &=&- f(\Omega_F,r,\theta)(B^{p})^2+
(B^{\hat{\phi}})^2,\label{beff1} \ea where $B^p$ is the poloidal
component of the magnetic field and \ba f(\Omega_F,r,\theta)
&\equiv& \Omega_F^2g_{\phi\phi}+2\Omega_F g_{0\phi}+g_{00}
\nonumber
\\ &=& -\left[\frac{\alpha^2 - \varpi^2(\Omega_F +
\beta)^2}{\alpha^2}\right]. \label{f}\ea  As a simple example,
consider a region very near to the horizon, $\alpha \rightarrow
0$.  Then Eq.(\ref{beff1}) can be written as \ba
 B^2-E^2
 &=&-\left[\frac{\varpi^2(\Omega_F -
\Omega_H)^2}{\alpha^2}\right](B^{p})^2+ (B^{\hat{\phi}})^2. \ea
One can see that without the toroidal component of the magnetic
field, $B^{\hat{\phi}}$, the above equation becomes negative and
the force-free condition breaks down.

The force-free condition is assumed for the plasma with a strong
magnetic field, which is sufficiently tenuous that the plasma
exerts no force on the magnetic field. If the whole region is
force-free, there is no place to extract energy  for the Poynting
flux.  We suppose a Poynting flux as a viable route to transport
the energy and angular momentum along the magnetic field lines.
Then there should be a region where matters exerts forces on the
magnetic field such that they convey the energy and angular
momentum along the field lines in the form of Poynting flux.  For
the Poynting flux out of accretion disk, it is naturally expected
that the force-free condition holds only for the magnetosphere not
on the disk.

\section{Basic Features of Poynting Flux}

To discuss the basic feature of the  Poynting flux, suppose an
artificial  surface at large radius which  encloses the  black
hole and accretion disk for example. The electromagnetic field
near the surface is described by the the Maxwell equation, \ba
 F^{\mu\nu}_{;\nu} &=&
4\pi J^{\mu}. \ea  The conservation of bulk current density should
be modified  since the bulk current density has a discontinuity on
this arbitrary  surface\cite{damour}.  To ensure the conservation
of the current  with this boundary,  the total conserved current
density $\mathcal{J}^\mu$ should be the sum of the bulk current
density and the surface current density $j^{\mu}$: \ba
\mathcal{J}^\mu=J^\mu + j^\mu, \label{jj} \ea where the  effective
surface current density are determined using the current
conservation law \ba
 \mathcal{J}^\mu_{\;\;;\mu}&=& 0. \label{current3}
\ea

As an  example,  let us consider  a spherical surface  for an
axisymmetric and steady case.  One can have \ba \sigma =
\frac{1}{4\pi}E^r, ~~ j^{\theta} = -\frac{1}{4\pi}B^{\phi}, ~~
j^{\phi} =\frac{1}{4\pi} B^{\theta}. \label{jsurface}\ea  If the
radius of the arbitrary surface is  located at far enough distance
from the central object, then we can use a flat space-time
geometry. Then the energy flux of electromagnetic field is given
by \ba {\cal E}^{r}=\frac{1}{4\pi} E^{\theta}B^{\phi},
\label{eflux}\ea which is the radial component of the Poynting
flux. On the arbitrary surface using Eq.({\ref{jsurface}), the
Poynting flux can be written as \ba {\cal
E}^{r}=-j^{\theta}E^{\theta}. \ea One can note that for a
non-vanishing Poynting flux  the force-free condition cannot be
maintained on the boundary, $ \vec{j}\cdot \vec{E}\neq 0$. It
corresponds to the `effective battery if it is positive (or 'ohmic
dissipator' if negative). Similarly the z-component of the angular
momentum flux can be written by \ba -\varpi B^{\phi}B^{r} =
-\frac{1}{4\pi}\varpi j^{\theta}B^r, \ea where Eq.(\ref{jsurface})
is used and $\varpi$ is a cylindrical radius. It corresponds to
the effective `torque' in $z$-direction exerted by the surface
current  although it is force-free in the original setting. This
observation  implies that the non-force-free nature of the system
sitting at the center inside the surface can be manifested on the
arbitrary boundary surface outside.

Now put the arbitrary surface  near to the gravitating object with
the  smallest radius but big enough to enclose the energy source
while keeping  outside to be  force-free: \ba F_{\mu\nu} J^{\nu}
=0. \ea
 We need a fully relativistic treatment.   For Kerr
geometry, the energy flux\cite{hklee} is given by \ba {\cal
E}^{\hat{r}}=\frac{\alpha}{4\pi} E^{\hat{\theta}}B^{\hat{\phi}} +
\frac{ \beta \varpi}{4\pi}B^{\hat{\phi}}B^{r}.
\label{fluxkerr1}\ea Using Eq.(\ref{fff2}), for a rigidly rotating
magnetic field line with angular velocity $\Omega_F$ in a
force-free environment up to the boundary surface,
Eq.(\ref{fluxkerr1}) can be written by \ba {\cal E}^{\hat{r}} &=&
\frac{\alpha}{4\pi} \frac{\Omega_F}{(\Omega_F
+\beta)}E^{\hat{\theta}}B^{\hat{\phi}}, \label{fluxkerr3}\ea which
shows a typical form of Poynting flux.

Using the effective currents defined above we can obtain \ba {\cal
E}^{\hat{r}} =\alpha j^{\theta} E^{\hat{\theta}} + \beta \varpi
j^{\hat{\theta}}B^{\hat{r}}. \label{fluxkerr2}\ea We can note that
the non-force-free nature of the source inside for the Poynting
flux outside is represented by non-vanishing $j^{\hat{\theta}}$
which `flows' across the magnetic field lines on the boundary
surface. Basically if the whole region is force-free then there is
no way to get a Poynting flux: no energy source. Hence the
magnetic field lines should be anchored onto a non-force-free
region from which the Poynting flux can carry out   the energy.
However the detailed nature of non-force-free region can only be
understood in the frame work of the relativistic
magnetohydrodynamics.

\section{Poynting Flux Dominated Accretion Flow}

We consider a system of a rotating black hole and a
two-dimensional accretion disk, which is assumed to be on the
equatorial plane in the background metric of a Kerr black hole and
the accretion is supposed to be driven by the Poynting
flux\cite{hkl6}

The stress-energy tensor is decomposed into two parts, the matter
part, $T^{\mu\nu}_D$, and the electromagnetic part,
$T^{\mu\nu}_{EM}$: \ba T^{\mu\nu}=T^{\mu\nu}_D +T^{\mu\nu}_{EM}
.\ea  In general the stress-energy tensor for the accretion disk
is determined by mass density, internal energy, pressure,
viscosity, radiative transfer and etc. \cite{acgl}. Since we are
interested in the accretion dominated by the Poynting flux, the
two-dimensional accretion disk is assumed to be non-viscous, cool,
and non-radiative such that
 the matter
part is given by \ba T^{\mu\nu}_D =\rho_m u^\mu u^\nu
\label{matter},\ea where $\rho_m$ is the rest-mass density and
$u^\mu$ is the four velocity of the accreting matter.
 It is also assumed
that there is a negligible mass flow in the direction
perpendicular to the disk: $u^{\theta} =0$.  We can obtain the
energy flux and the angular momentum flux given by
\begin{eqnarray}
{\cal E}^{\mu}_{D} = -\rho_m u_0 u^{\mu}, ~~ {\cal L}^{\mu}_{D} =
\rho_m u_{\phi} u^{\mu}. \label{Lm}\end{eqnarray}

For an idealized thin disk on the two-dimensional plane we take
\begin{eqnarray}
\rho_m = \frac{\sigma_m}{\rho} \delta(\theta-\pi/2),
\end{eqnarray} where $\sigma_m$ is the surface rest-mass density.
The rate of the rest-mass flow crossing the circle  of radius $r$
defines
 the mass accretion rate $\dot{M}_+$  by
\begin{eqnarray}
\dot{M}_+ = - 2\pi \sigma_m \rho u^r \label{M+0}.\end{eqnarray} It
 is identical to the one  derived in \cite{acgl} and \cite{gp}, in
which the vertical structures of the disk  are integrated. The
stationary accretion flow implies that  $\dot{M}_+$ is
$r$-independent. Using Eq.(\ref{Lm}) the radial flow of the energy
and angular momentum of the matter at $r$  can be given by $ u_0
\dot{M}_+ ~~ {\rm and}~~ - u_{\phi}~ \dot{M}_+ $ respectively.

The horizon of a black hole is a mathematically well defined
surface, on which  the appropriate surface current density has
been studied in depth\cite{TPM}.  On the other hand, the surface
of the physical accretion disk is not expected to have any sharp
boundary and various  shapes have been suggested depending on the
properties of the accretion flow. In this work, however,  we
assume a simplified accretion disk, which is vertically squeezed
down to the equatorial plane, a two-dimensional accretion disk.
Then we can associate the surface current density in Eq.(\ref{jj})
in a simple way,
 \ba j^\mu={1\over
        4\pi}(F^{\theta\mu}_{+}-F^{\theta\mu}_{-})
        \delta(\theta-\pi/2), \label{surface-current} \ea
which are nothing but Gauss' law and Ampere's law as given by \ba
\sigma_e = -{1\over
        2\pi}E^{\hat{\theta}}_+ , ~~
        K^{\hat{r}} =-{1\over
        2\pi}B^{\hat{\phi}}_{+} ,\quad
        K^{\hat{\phi}}={1\over 2\pi}B^{\hat{r}}_{+}
        \label{bc2s},\ea
where $\sigma_e$ and $K^i$ are surface charge and surface current
density(spatial) respectively  and  the reflection symmetry with
respect to the equatorial plane,
$F^{\theta\mu}_{+}=-F^{\theta\mu}_{-}$, is used.  The  surface
current densities are responsible for the Poynting flux  from the
disk.  Hence one can consider the singular equatorial plane with
non-vanishing surface currents manifests the non force-free nature
of the accretion flow driven by Poynting flux.

The energy and the  angular momentum of the disk can be carried
out by Poynting flux  along the magnetic field lines which are
anchored on the disk. In our simplified  model, it is the main
driving force for the accretion flow.

Using the Killing vector in $t$-direction, $\xi^{\mu} = (1,\, 0,\,
0,\,0),$ we can define the energy flux ${\cal E}^{\mu}$ from the
stress-energy  tensor of the electromagnetic field,
$T^{\mu\nu}_{EM}$,
\begin{eqnarray}
{\cal E}^{\mu}_{EM} = - T^{\mu\nu}_{EM}\xi_{\nu} ,
\end{eqnarray} where
\begin{eqnarray}
T^{\mu\nu}_{EM} = \frac{1}{4\pi}(F^{\mu}_{~ \rho}F^{\nu\rho}
-\frac{1}{4}
g^{\mu\nu}F_{\rho\sigma}F^{\rho\sigma}).\label{tmunu1}
 \end{eqnarray}

 For the
 steady and axisymmetric case in this work, $E^{\phi}=0$, and we
 get the total energy flux of electromagnetic field in
 $\theta$-direction, ${\cal E}^{\theta}_{D}$, given by
   \ba {\cal E}^{\theta}_{D} &=&
\frac{1}{\rho}[\alpha K^{\hat{r}}E^{\hat{r}}_{+}  - \beta \varpi
 K^{\hat{r}}B^{\hat{\theta}}_{+}]. \label{erhatd2}
\end{eqnarray}
The  first term in Eq.(\ref{erhatd2}) can be considered to be an
ohmic interaction, $\vec{E}\cdot \vec{K}$. For the current in the
direction of the  electric field tangential to  the disk there
might be  an  energy dissipation into the disk surface. This is
what one can expect on the black hole horizon\cite{TPM}. However
for the current in the opposite direction this term corresponds to
the electro-motive force and it is the case for the accretion disk
on the equatorial plane discussed by Blandford and
Znajek\cite{blandford,BZ}as well as in this work. The second term
can be interpreted as  a magnetic braking power on the rotating
body with an angular velocity $\omega = - \beta$.

 Using the Killing vector in
$\phi$-direction for the axial symmetric case, $\eta^{\mu} = (0,\,
0,\, 0,\,1)$, we can get also the total flux  of the angular
momentum in $\theta$-direction, ${\cal L}^{\theta}_{D}$,
\begin{eqnarray}
{\cal L}^{\theta}_D = -\frac{\varpi}{2\pi \rho}B^{\hat{\theta}}_+
B^{\hat{\phi}}_+\label{ltheta} = \frac{1}{\rho}\varpi K^{\hat{r}}
B^{\hat{\theta}}_+,  \end{eqnarray} which is nothing but a
magnetic torque exerted on the surface current density $K^{r}$.

Now the Poynting power measured at infinity is given  by \ba P_{D}
&=& -\int dr r (-\alpha E^{\hat{r}}_+ B^{\hat{\phi}}_+ + \beta
\varpi B^{\hat{\phi}}_+B^{\hat{\theta}}_+), \label{pd} \ea and the
rate of angular momentum transfer, $P^L_{D}$,  out of the
accretion flow is given by \ba P^L_{D} &=&\int dr r \varpi
B^{\hat{\phi}}_{+}B^{\hat{\theta}}_{+}. \label{plaf} \ea

\subsection{Accretion Equations and Stream Equation}

The dynamics of an accretion flow is determined  by the
conservation equation of the stress-energy tensor:
 \ba T^{\mu\nu}_{\quad;\mu}=0.\label{cons}\ea
From the  conservation of stress-energy tensor,
 one can
obtain the accretion equations \cite{hklee2} driven by the
Poynting flux in a force-free magnetoshpere given by
 \ba
(\partial_r u_0) \m +r\w \Omega_F B_+^{\hat{\phi}}
B_+^{\hat{\theta}}=0
        \label{tc4} ,\\
        (\partial_r u_\phi) \m -r\w
        B_+^{\hat{\phi}}B_+^{\hat{\theta}}=0 \label{tc5} ,\\
   {\m\over 2\pi r^2} g^{rr} {u^0\over u^r}(\partial_r u_\phi)
        \left[ \Omega_D+{\partial_r u_0 \over
        \partial_r u_\phi} \right] \nonumber \\-{\sqrt{\Delta}\over 2\pi
        r^2}B_+^{\hat{r}}B_+^{\hat{\theta}}\left({\w^2\over\alpha^2}
        \Omega_F^2-1 \right) =0  \label{tc6} ,\ea
where the angular velocity of the disk is given by $ \Omega_D=
u^\phi/u^0$.  Eqs.(\ref{tc4}) and (\ref{tc5}) correspond to the
energy and angular momentum conservation respectively in the
stationary and axisymmetric setting in this work. In
Eq.(\ref{tc4}) the radial change of the energy flux of accreting
matter, $-u_0 \dot{M}_+$,  is balanced by the Poynting flux in
theta direction\cite{hklee}, ${\cal E}^{\theta}_{D}$. Similarly
the radial change of the angular momentum flux, $u_{\phi}
\dot{M}_+$ is balanced by ${\cal L}^{\theta}_{D}$.   The radial
equation Eq.(\ref{tc6}) essentially determines the orbital motion
of the disk.  In the absence of external fields in Eq.(\ref{tc6}),
the angular velocity is determined as \ba \Omega_D =-{\partial_r
u_0 \over
\partial_r u_\phi},  \ea which is one of the characteristics of
the Keplerian orbit.  Hence deviations of $\Omega_D$ from the
Keplerian one  are naturally  expected in the Poynting flux
dominated accretion disk.

It should be noted the force-free condition holds only for the
magnetosphere not on the disk. In fact  the driving force for an
accretion flow on the disk is the magnetic braking in
Eqs.(\ref{tc4}) and(\ref{tc5}). Then Eq.(\ref{pd}) becomes  \ba
P_{D} &=& \int dr r ( \varpi \Omega_F B_+^{\hat{\theta}}
B^{\hat{\phi}}_+). \label{pergoff} \ea

From Eqs.(\ref{tc4}) and (\ref{tc5}), one obtains an interesting
relation \ba {\partial_r u_0\over\partial_r u_\phi}=-\Omega_F
\label{omegaf}.\ea which has no explicit dependence on the field
configuration. It seems to imply that $\Omega_F$ is determined
essentially by the dynamics of the disk. However one should note
that the dynamics itself is governed not only by the gravity but
also by the electromagnetic field as well.

The configuration of the ordered magnetic field around the
accretion disk  and the Poynting outflow from the disk has been
discussed both in analytical and numerical
studies\cite{ga,lop,fendt,love}. For the strong enough
electromagnetic field around the compact object, the force-free
magnetosphere can be established. In the non-relativistic
formulation, Blandford\cite{blandford} suggested an axisymmetric
and stationary electromagnetic field configuration around  an
accretion disk on the equatorial plane. The poloidal field
configuration for a black hole is known to satisfy a second order
elliptic differential equation called Grad-Shafranov
equation\cite{beskin} or a stream equation\cite{MT} for the stream
function $\Psi$ and current $I$. The poloidal and toroidal
components of the  magnetic field can be written in terms of
$\Psi$ and $I$ respectively:
 \ba \vec{B}^P = {1\over2\pi\w}\nabla \Psi \times e^{\hat{\phi}} ,
\quad B^{\hat{\phi}} = -{2I\over \w \alpha} \label{bpt}.   \ea
 Possible types of solutions in a  force-free
magnetosphere has been discussed recently in the relativistic
formulation\cite{ghosh,llk}.  To calculate the Poynting flux  we
need to know the configuration of the magnetic field which is
consistent with the accretion equations.  And the problem is
reduced  to solve the four coupled equations, three accretion
equations and a Grad-Shafranov equation(stream equation).

\subsection{Example: Numerical Solutions in Schwarzschild Background}

The coupled differential equations in a Kerr geometry in the
previous section is not easy  to solve.  To show some examples of
the solutions  we try a numerical solution\cite{jp} in a simpler
background, Schwarzschild background.  In the Schwarzschild
background, the stream equation is given by
        \ba  & & \partial_r \left\{\left(1-{2M\over r}\right)\partial_r \Psi
        \right\}+{\sin\theta\over r^2}
        \partial_\theta\left({1\over\sin\theta}\partial_\theta \Psi \right)
        \nonumber \\&-&\Omega_F \sin^2\theta\partial_r \left(r^2 \Omega_F \partial_r
        \Psi \right)
        -{\Omega_F\over 1-{2M\over
        r}}\sin\theta\partial_\theta\left(\sin\theta\Omega_F\partial_\theta\Psi\right)
        \nonumber \\&=& -{16\pi^2
        I {dI\over d\Psi}\over \left(1-{2M\over r}\right)}\label{st2}.\ea
In general there is no known analytic forms for the solutions. In
the limiting case when $\Omega_F$ and $I$ vanishes, one of the
solutions, $\Psi_0$, suggested by Blandford and Znajek \cite{BZ}
is given by
        \ba \Psi_0 =\pi C X
        \label{psi0},\ea
where \ba X\equiv r(1 \mp \cos\theta)+2M
        (1 \pm \cos\theta)\{1-\log(1 \pm \cos\theta)\}. \ea
For non-vanishing $\Omega_F$ and $I$, we consider a solution for
which the shape of the magnetic surface are the same as the
magnetic surface defined by $\Psi_0$\cite{m84}. Thus $\Psi , I$
and $\Omega_F $ are assumed to depend  on X. We suppose that the
derivative of $\Psi$ has the same form as in the flat background,
Eq.(\ref{psi0}). That is, we take an Ansatz such that
        \ba {d\Psi\over d X}={\pi C \over(1+\Omega^2_F
        X^2)^{1/2}}.
        \label{ansatz}\ea

Four basic equations governing the accretion flow under the
influence of paraboloidal-type configuration  are solved
numerically\cite{jp}. The radial variations of $u_0$, $u_{\phi}$,
$I$ and $\Omega_F$ on the disk are  basically   functions of the
accretion rate and  the strength of the magnetic field. Numerical
calculations show  that the angular velocities  of the magnetic
field lines($\Omega_F$) are  different from either the Keplerian
angular velocity($\Omega_K$) or the disk angular
velocity($\Omega_D)$. It implies that the two-dimensional
approximation of the perfect conducting disk, for which  $\Omega_F
= \Omega_D$,   may not be implemented particularly with the
paraboloidal type configuration. As expected in the previous
section for the velocity of the magnetic field line
$v_F(=r\Omega_F)$ less than the  speed of light, $\Omega_F$ is
found to be larger than $\Omega_D$.

It is also found that the strength of the magnetic field is
increasing as $r$ goes near the inner edge.   Since the magnetic
field  as well as the surface current are increasing as $r$ gets
smaller, it is naturally expected that the Poynting flux increases
substantially as $r$ approaches to the center . Although the
numerical calculation does not go beyond $r<6M$, it may indicate a
possible electromagnetic jet structure  near the inner edge of the
disk.

\section{Poynting Flux from Ergosphere}
On the horizon inside the ergosphere,  we can make use of the
Znajek's boundary condition\cite{znajek},
 \ba B^{\hat{\phi}} = \frac{\varpi(\Omega_F-\Omega_H)}{\alpha}
B^{\hat{r}}, \ea  and we get \ba B^2-E^2
 =-\left[\frac{\varpi^2(\Omega_F -
\Omega_H)^2}{\alpha^2}\right](B^{\hat{\theta}})^2. \label{beffh}
\ea  It implies that  for the magnetosphere with $B^{\theta}=0$ on
the horizon the force-free  nature can be justified and   the
region where the electromagnetic field lines can exert non-zero
force should be inside horizon. As discussed in section III, we
can define a non-zero surface current on the horizon boundary such
that the rotation of black hole can be slowed down or speeded up
and we get the Poynting flux\cite{BZ}.  Using
Eq.(\ref{fluxkerr3}), the rate of energy extraction at infinity
can be evaluated using the electromagnetic field on the horizon,
\ba P_{hole}=-\int_{r=r_H}
 \frac{1}{4\pi}\Omega_F(\Omega_F-\Omega_H )
\varpi^2(B^{\hat{r}})^2 d\theta d\phi.\label{bzrate} \ea It is to
be noted that the force-free condition might not be maintained on
the horizon  for  $B^{\hat{\theta}}\neq 0$. It may corresponds to
the case when  the force-free condition is invalid in the vicinity
of horizon\cite{punsly}, which,  however, depends on the details
of the physical properties of the plasma.

We will explore whether similar analysis  can be applied to the
magnetic field lines threading on the equatorial plane inside the
ergosphere\cite{lb}. The overall environment and physical
processes are assumed to be axisymmetric and steady in the
background of Kerr geometry. The effects on the geometry due to
the energy density of the electromagnetic field and the accreting
material are assumed to be negligible. In this work we assume a
very thin accretion flow on the equatorial plane. Let us suppose
the force-free relation, $F^{\mu\nu}J_{\nu} =0$, holds up to the
equatorial plane to see under what condition  $B^2 - E^2
>0$ is violated. Because of the reflection symmetry in the equatorial plane,
there is no current crossing the equatorial plane and $B^{\phi}
=0$ on the equatorial plane.  Then we get \ba B^2-E^2 &=&
-f(\Omega_F,r,\theta=\pi/2)\frac{(B^{p})^2}{ \alpha^2},
 \ea
where $f=0$ determines the light surfaces\cite{takahashi}. For a
given $r, $ the sign of $B^2-E^2$ is then determined by $f$. Hence
the condition for force-free requirement , $f<0$, determines the
acceptable range of the angular velocity, $\Omega_F(r)$, of the
field lines which cross  the equatorial plane at $r$.  One obtains
\ba \Omega_{-}(r) \leq \Omega_F(r) \leq \Omega_{+}(r), \ea where
\ba \Omega{\pm} = \Omega_{ZAMO} \pm \frac{\alpha}{ \varpi}, ~~
\Omega_{ZAMO}= -\beta. \ea For such magnetic field lines which
rotates with $\Omega_{-}(r) \leq \Omega(r) \leq \Omega_{+}(r)$, we
cannot expect any Poynting flux since it is simply force free. In
other words, we cannot get non-vanishing $B^{\hat{\phi}}$  for the
Poynting flux. It is because for the force-free magnetosphere,
$B_T = \varpi \alpha B^{\hat{\phi}}$ is constant on the magnetic
surface and therefore $B^{\hat{\phi}}$ should remain to be zero
not only on the equatorial plane but also off equatorial plane
along the  field line.

However for the field line with  the angular velocity smaller than
$\Omega_- $,  the region on the equatorial plane cannot be
consistent with the force-free requirement. Hence the magnetic
field line with $\Omega_F < \Omega_-$ can exert a force on the
non-force-free region near  the plane. And we can get a Poynting
flux along such field lines, which pass through the non-force-free
region. Consider a  non force-free region with finite width around
equatorial plane.  Since it is not a force-free region, the
toroidal component can be developed along the field line off the
plane,  although on the equator it should be zero. It reaches the
boundary of force-free region with $B^{\phi}=B^{\phi}_+$, which is
determined by \ba -f(\Omega,r,\theta)\frac{(B^p_+)^2}{ \alpha^2} +
(B^{\hat{\phi}}_+)^2 =0. \ea  We can guess $\Omega_F$ is not much
different form $\Omega_-$:\ba \Omega_F = \Omega_{-} + \delta, \ea
where $\delta$ is determined by $B^{\phi}_+$.  It also implies
that the Poynting flux out of equator is  possible only inside the
ergosphere since $\Omega_- =0$ on static limit and outside the
ergosphere there is no Poynting flux.  It is consistent with the
recent numerical simulations by Komissarov\cite{kom1}.

Let us consider a simple example where  the non force-free region
can be approximated as a thin disk, two dimensional disk on the
equatorial plane,  dominated by the inertia. Then we can suppose
no-vanishing $B^{\hat{\phi}}$ up to the equatorial plane with
discontinuity at the plane. In this boundary value problem,   the
surface charge,  $\sigma_e$, and the current density, $K^i$, can
be defined on the equatorial plane as in the previous section. The
energy flux in $\theta$-direction is given by Eq.(\ref{erhatd2}).
 Inside the disk, it is not
force-free and there is no reason that $B_T =\varpi\alpha B^{\phi}
$ should be conserved along the field lines and $B_T$ or
equivalently $B^{\phi}$ develops from zero to finite value up to
the disk surface.  Only beyond the surface where force-free
condition can be realized, $B_T$ is conserved along the magnetic
surface. The Poynting power measured at infinity is given  by \ba
P_{ergo} &=& \int^{r_o}_{r_H} dr r \varpi \Omega_F
B^{\hat{\phi}}_{+}B^{\hat{\theta}}_{+}. \label{pergo} \ea

It is interesting to compare the power from the disk inside the
ergosphere, $P_{ergo}$,  to the Poynting power from  a black hole
$P_{hole}$, Eq.(\ref{bzrate}). As an example we take a
magnetosphere similar to that suggested by
Blandford\cite{blandford}. Using a set of approximations for the
magnetic field and $\Omega_F$ for a numerical estimation\cite{lb},
it is found that
  the ratio of the Poynting power from the
ergosphere to that from the black hole,
$P_{ergo}(\tilde{a})/P_{hole}(\tilde{a})$, is increasing as the
angular momentum parameter $\tilde{a}$  increases. The ratio of
the angular momentum flux,
$P^{L_{\phi}}_{ergo}(\tilde{a})/P^{L_{\phi}}_{hole}(\tilde{a})$,
is also increasing with $\tilde{a}$.  In this simple analysis it
is observed  that for a maximally rotating black hole the power
from the disk inside the ergosphere can be  as much as $30 \%$ of
the power from the black hole.

Physically $P_{ergo}$ is  a part of the gravitational energy of
the particles of the negative energy orbit in  a disk on the
equatorial plane tapped by the magnetic field.   For example, if
the energy momentum tensor of the two dimensional disk is
dominated by the inertia, mass density $\rho_m$, we can make use
of Eqs.(\ref{tc4}) and (\ref{tc5}). We can see that the second
terms in these equations correspond to $P_{ergo}$ and
$P^{L_{\phi}}_{ergo}$ which depend on the energy(negative) of the
orbit and the mass accretion rate. The estimation of  the first
terms in Eqs.(\ref{tc4}) and (\ref{tc5}) inside the ergosphere is
an interesting subject to be studied in the future to get a more
realistic estimation of $P_{ergo}$.

\section*{Acknowledgment}
\noindent The author  would like to thank  Roger Blandford for
discussions and the  kind hospitality during his stay at KIPAC. He
also thanks Dmitri Uzdensky for valuable comments.  This  was
supported by grant No. (R01-2006-000-10651-0) from the Basic
Research Program of the Korea Science \& Engineering Foundation.

\end{document}